\documentclass[12pt]{article}   
\usepackage{amsmath}

\title{Quadrupoles, to Third Order}
\author{Rick Baartman, TRIUMF}
\date{March, 2014}
\begin{document}
\maketitle
\section{Introduction}
I derive the third order optics of quadrupoles. I transform away the derivatives of the strength function, and thus demonstrate that third order aberrations are insensitive to fringe field shaping. The results can be used for efficient tracking to third order or for simple and quick evaluation of nonlinear effects.

This is an extension to my work of 1997\cite{Baartman:1997zza}, which covered both electrostatic and magnetic quads, but only for the non-relativistic case.

\section{Theory}
A particle of charge $q$ and mass $m$ has canonical pairs of coordinates $((x,P_x),(y,P_y),(z,P_z),(t,E))$. The equation coupling these is
\begin{equation}
 (E-q\Phi)^2-|\vec{P}-q\vec{A}|^2c^2=m^2c^4.
\end{equation} 
Here, $\Phi$ and $\vec{A}$ are respectively the scalar and vector potentials; both are functions of $(x,y,z)$, and both are zero on the reference particle's orbit.

Let us use units for time, energy and potential that sets resp.\ $c=1$, $m=1$, $q=1$. This is permissible as long as we consider no processes that change mass or charge.
\begin{equation}
 (E-\Phi)^2-|\vec{P}-\vec{A}|^2=1.
\end{equation} 
To find the Hamiltonian $H=E$, solve for $E$: 
\begin{equation}
H(x,P_x,y,P_y,z,P_z;t)=\Phi+\sqrt{1+|\vec{P}-\vec{A}|^2}
\end{equation} 

\subsection{In Frenet-Serret Frame}
The reference orbit is assumed to be in one plane and is generally curved with curvature $h=1/\rho$. The transformation is conventionally made to the Frenet-Serret coordinate system where the longitudinal coordinate $s$ is in the reference orbit direction, so $h=h(s)$, $x$ is radially outward, and $y$ is perpendicular to the bend plane. Ruth\cite{ruth1986single} shows that then the Hamiltonian is
\begin{equation}
 H(x,P_x,y,P_y,s,P_s;t)=\Phi+\sqrt{1+(P_x-A_x)^2+(P_y-A_y)^2+\frac{(P_s-A_s)^2}{1+hx}}
\end{equation} 

As conventional in beam and accelerator physics, we use the longitudinal coordinate $s$ as independent variable. Then the Hamiltonian is $-P_s$:
\begin{equation}
 H(x,P_x,y,P_y,t,E;s)=-A_s-(1+hx)\sqrt{(E-\Phi)^2-1-(P_x-A_x)^2-(P_y-A_y)^2}
\end{equation} 

\subsection{In Differential Coordinates}
The ``reference particle'' has $x=y=0$ and $P_x=P_y=0$. In the following, we use the traditional symbols $\beta$, $\gamma$ and hence also $\beta\gamma$ for the reference particle's speed, energy, and momentum, respectively.

The Hamiltonian is awkward because it mixes small dynamic quantities $x,y,P_x,P_y$ with a large one $E$.
We only care about particles with a small $\Delta E$ deviation from the reference energy $\gamma$, and a small $\Delta t$ deviation from the reference time $t_0=s/\beta$. We do this with a canonical transformation from $(t,-E)$ to $(\Delta t,-\Delta E)$. The generating function is \begin{equation}\label{gend}
F(t,-\Delta E)=\left(t-\frac{s}{\beta}\right)(-\Delta E-\gamma)
\end{equation} 
The new Hamiltonian is 
\begin{equation}\label{newh}
\tilde{H}_s=H_s+\frac{\partial F}{\partial s}=H_s+\frac{\gamma+\Delta E}{\beta}
\end{equation} 
Furthermore, we introduce new coordinates $(\tau,P_\tau)$ in place of $(\Delta t,\Delta E)$, with $\tau=\beta\Delta t$, $P_\tau=\Delta E/\beta$. This results in a new ``time'' coordinate $\tau$ being the distance ahead of the reference particle, and the ``energy'' coordinate being the momentum deviation w.r.t.\ the reference particle.

The Hamiltonian is then
\begin{equation}\begin{split}
 &\tilde{H}_s(x,P_x,y,P_y,\tau,P_\tau;s)=\\
&P_\tau-A_s-(1+hx)\sqrt{(\gamma+\beta P_\tau-\Phi)^2-1-(P_x-A_x)^2-(P_y-A_y)^2}
\end{split}\end{equation} 

\subsection{Re-Normalize Momenta}
We now change the units of momentum to $\beta\gamma$. This will have the advantage that outside of the regions of electric and magnetic fields, we have $P_x=x'$, $P_y=y'$, $P_\tau=\Delta P/P$ where primes are derivatives w.r.t.\ $s$. Further, we rescale scalar potential by a factor $\beta^2\gamma$, and vector potential by a factor $\beta\gamma$. The result is
\begin{equation}
 H=P_\tau-A_s-(1+hx)\sqrt{\left(\frac{1}{\beta}+\beta(P_\tau-\Phi)\right)^2-\frac{1}{\beta^2\gamma^2}-(P_x-A_x)^2-(P_y-A_y)^2}
\end{equation} or
\begin{equation}
 H=P_\tau-A_s-(1+hx)\sqrt{1+2(P_\tau-\Phi)+\beta^2(P_\tau-\Phi)^2-(P_x-A_x)^2-(P_y-A_y)^2}
\end{equation} 
This is the general Hamiltonian. It is exact.

\subsection{Potentials' Scales}
For handy reference, here are the definitions of the scaled potentials in terms of the unscaled (subscript u):
\begin{equation}\label{scale}\begin{split}
\Phi&=\frac{q}{\beta^2\gamma mc^2}\Phi_{\rm u}(x,y,s),\\
\vec{A}&=\frac{q}{\beta\gamma mc}\vec{A}_{\rm u}(x,y,s)
\end{split}\end{equation} 

\subsection{Relativistic Limits}
This $H$ also has the nice feature that the non-relativistic and ultra-relativistic limits are simple:
\begin{equation}\begin{split}
 \beta\ll 1 &: H=P_\tau-A_s-(1+hx)\sqrt{1+2(P_\tau-\Phi)-(P_x-A_x)^2-(P_y-A_y)^2}\\
\gamma\gg 1 &: H=P_\tau-A_s-(1+hx)\sqrt{(1+P_\tau-\Phi)^2-(P_x-A_x)^2-(P_y-A_y)^2}
\end{split} \end{equation} 

\subsection{Example: Field-free, curvature-free}

\begin{equation}
\tilde{H}_s=P_\tau-\sqrt{1+2P_\tau+\beta^2P_\tau^2-P_x^2-P_y^2}
\end{equation} 
Then $P_x'=P_y'=P_\tau'=0$, and to first order:
\begin{equation}
 x'=\frac{\partial H}{\partial P_x}=\frac{P_x}{\sqrt{1+2P_\tau+\beta^2P_\tau^2-P_x^2-P_y^2}}\approx P_x
\end{equation}
and similar for $y'$, 
\begin{equation}
 \tau'=1-\frac{1+\beta^2P_\tau}{\sqrt{1+2P_\tau+\beta^2P_\tau^2-P_x^2-P_y^2}}\approx \frac{P_\tau}{\gamma^2}
\end{equation} 

\subsection{Straight elements, ignore longitudinal}

Let us now confine ourselves to straight elements ($h=0$) and concentrate only on transverse. Then for magnetic elements, we have
\begin{equation}\label{HmagQ}
   H=-A_s-\sqrt{1-(P_x-A_x)^2-(P_y-A_y)^2}
\end{equation}
and for electrostatic elements, we have
\begin{equation}\label{HelQ}
   H=-\sqrt{1-2\Phi+\beta^2\Phi^2-P_x^2-P_y^2}
\end{equation}

\section{Electrostatic Quads} 
Compared with equation 2 of the 1997 paper\cite{Baartman:1997zza}, we notice an extra term $\beta^2\Phi^2$ in the square root of eqn.\,\ref{HelQ}\footnote{There is a also a factor of 2 because the scaling of eqn.\,\ref{scale} differs from the 1997 scaling by this factor}.

We expand the square root to 4${\rm th}$ order in coordinates and ignore the constant:
\begin{equation}
  \label{H2}
  H\approx {1\over 2}(2\Phi-\beta^2\Phi^2+P_x^2+P_y^2)+{1\over 8}(2\Phi+P_x^2+P_y^2)^2.
\end{equation}

To the same order, Laplace's equation gives for the expansion of the quadrupole potential: 
\begin{equation}
  \label{V1}
  \Phi={k(s)\over 2}(x^2-y^2)-{k''(s)\over 24}(x^4-y^4).
\end{equation}
The expanded Hamiltonian, correct to 4$^{\rm th}$ order is
\begin{equation}
  \label{H3}
\begin{split}
H&={k(x^2-y^2)\over 2}+{P_x^2\over 2}+{P_y^2\over 2}+\\
&+{(P_x^2+P_y^2)^2\over 8}+{k(x^2-y^2)(P_x^2+P_y^2)\over 4}+\\
&+{k^2(x^2-y^2)^2\over 8\gamma^2}-{k''(x^4-y^4)\over 24}
\end{split}
\end{equation}

\subsection{$k''$ is a ``Fringe Field Effect''?}
The trouble with applying this to simple cases like thin lenses and
hard-edge limits is the presence of $k''(s)$, which becomes singular in
those limits. In most cases, one sacrifices physical insight and simply
traces particles with this Hamiltonian, using a more-or-less realistic
function $k(s)$. For example, the approach taken in {\tt GIOS}\cite{matsuda1972third} is
to leave it up to the user to specify `fringe field integrals' such as
$\int k^2ds$ through the fringe fields. However, this leaves one quite vulnerable to
error; different integrals may not be realistic or consistent with each
other.  Moreover, if one needs to solve Laplace's equation to find fringe
field integrals, one might as well use the solution directly in a
ray-tracing code. If one does go through this exercise, one discovers that
the higher order aberrations are relatively insensitive to the `hardness'
of the quadrupole edges. This leads one to suspect that the aberrations are
dominated by an intrinsic effect which has nothing to do with the detailed
shape of the fringing field. Such is indeed the case.

\subsection{$k''$ can be transformed out!}
It turns out to be possible to find a canonical transformation which
eliminates the derivatives of $k(s)$. In our case, we wish to
retain terms to 4$^{\rm th}$ order in the Hamiltonian (3$^{\rm rd}$ order
on force), and the transformation $(x,P_x,y,P_y)\rightarrow (X,P_X,Y,P_Y)$
has generating function
\begin{equation}
  \label{gen}
  G(x,P_X,y,P_Y)=xP_X+yP_Y+{k'\over 24}(x^4-y^4)-{k\over 6}(x^3P_X-y^3P_Y).
\end{equation}
To the same order, this yields the transformation
\begin{equation}
  \label{trae}
\begin{split}
  x&=X+{k\over6}X^3 \\
  P_x&=P_X-{k\over2}X^2P_X+{k'\over6}X^3.
\end{split}
\end{equation}
The $y$-transformation is obtained by replacing $x,P_x,X,P_X$ with
$y,P_y,Y,P_Y$ and $k$ with $-k$. Note that outside the quadrupole, the
transformed coordinates are the same as the original ones.

This yields the transformed Hamiltonian $H^\ast$:
\begin{equation}
  \label{Hstar}
\begin{split}
H^\ast&={k\over 2}(X^2-Y^2)+{1\over 2}(P_X^2+P_Y^2)+\\
&+{1\over 8}(P_X^2+P_Y^2)^2-{k\over 4}(X^2+Y^2)(P_X^2-P_Y^2)+\\
&+{(7-3\beta^2)k^2\over 24}(X^4+Y^4)-{(1-\beta^2)k^2\over 4}X^2Y^2.
\end{split}
\end{equation}
We can identify the terms: the first two are the usual linear ones; the
third term is not related to the electric field (it is small and due to the
fact that $x'\neq P_x$ or, equivalently, $\tan\theta\neq\sin\theta$); the
4$^{\rm th}$ term is also small and arises because a particle going through the
quadrupole at an angle is inside the quad for slightly longer than one
which remains on axis. See ref.\,\cite{baartman_9521} for more complete physical
derivation of the individual terms. 

\subsection{Thin lens, Hard Edge Formulae}
The dominating higher order terms are the last two terms in
eqn.\,\ref{Hstar}. Since there are no derivatives of $k$, we can directly
write down the aberrations in the thin-lens limit:
\begin{equation}
  \label{tlens}
  \Delta P_x={-1\over f^2L}\left({7-3\beta^2\over 6}x^3-{1-\beta^2\over 2}xy^2\right),
\end{equation}
with a similar expression for $\Delta P_y$.  $L$ and $f$ are the
quadrupole's length and focal length. (Actually, it is more accurate to replace ${1\over f^2L}$ with $\int k^2ds$.) 

The fractional focal error
is found by dividing by the linear part $\Delta_0P_x=-x/f$:
\begin{equation}
  \label{frace}
  {\Delta f_x\over f}={1\over fL}\left({7-3\beta^2\over 6}x^2-{1-\beta^2\over 2}y^2\right)
\end{equation}
for $x$, and similarly for $y$.

\subsection{Physical Interpretation: Speed Effect}
It is interesting and instructive to deconstruct the final result
\ref{tlens} to derive physical origins for these terms. We do this in
the thin lens limit.

Referring to the
untransformed Hamiltonian (\ref{H3}), we can identify the term with
$\gamma^{-2}$ as due to a ``velocity-gain'' effect: particles entering
the electric field have their speed changed because of longitudinal
field, for example, those entering near the like-charged electrode are
slowed so spend a longer than normal time in the focusing field. This
effect disappears in ultra-relativistic limit. The contribution to $P_x$
from this effect in thin lens limit is
\begin{equation}
  \label{tlensv}
  \Delta P_x|_{dv}= -{x(x^2-y^2)\over 2\gamma^2}\ \int k^2ds,
\end{equation}
leaving only ${2\over 3}x^3$ inside the parentheses of eqn.\,\ref{tlens} to account for.

\subsection{Physical Interpretation: $k''$ Effect}
The direct effect of the $k''$ term in the potential and the Hamiltonian
(\ref{H3}) can be found from
integrating by parts:
\begin{equation}
  \label{eq:yu}
  \begin{split}
    \Delta P_x|_{k''}&=-\int\left.{\partial H\over\partial x}\right|_{k''}ds={1\over 6}\int k''x^3ds=-{1\over 2}\int k'x^2x'ds\\ &\approx {1\over 2}\int kx^2x''ds\approx-{1\over 2}\int k^2x^3ds\approx-{x^3\over 2}\int k^2ds
  \end{split}
\end{equation}

\subsection{Physical Interpretation: $\Delta x$ Effect}
The remainder is now $-{x^3\over 6}\int k^2ds$. This originates from a small, subtle and largely overlooked effect; I overlooked it in my earlier work\cite{baartman_9521}. It originates from a shift in $x$ experienced in the quad. 

From \ref{eq:yu} above, we also find
\begin{equation}
  \label{eq:dx}
  x''={k''x^3\over 6}
\end{equation}
which in thin lens approx can be integrated directly to obtain
\begin{equation}
  \label{eq:Dx}
  \Delta x={kx^3\over 6}.
\end{equation}

Another way to see this is from eqn.\,\ref{trae}. Since the transformed variable $X$ does not see any shifts due to derivatives of $k$, it is unaffected on passing through the fringe field. But if $X$ is not shifted, then $x$ must be shifted by $kx^3/6$.

As $x$ is shifted, there results a different overall focus effect $\Delta P_x'=k\Delta x$:
\begin{equation}
  \label{eq:pop}
  \Delta P_x|_{dx}\approx -{x^3\over 6}\int k^2ds.
\end{equation}

Equations \ref{tlensv}, \ref{eq:yu}, \ref{eq:pop}, when summed, give \ref{tlens}. Q.E.D.

\section{Magnetic Quads, Scalar Potential} 
We can use the same scalar potential for magnetic as for electrostatic (\ref{V1}),
but rotated by $\pi/4$:
\begin{equation}
\Psi(x,y,s)=k(s)xy-{k''(s)\over 12}xy(x^2+y^2)
\end{equation}
To find the vector potential, we follow Venturini-Abell-Dragt\cite{venturini1998map} (VAD)
and express it first in polar coordinates:
\begin{equation}
  \label{eq:psipol}
  \Psi(r,\theta,s)=\left({k\over 2}r^2-{k''\over 24}r^4\right)\sin2\theta
\end{equation}
\subsection{Vector Potential, Gauge Choice}
But instead of VAD's gauge condition $A_\theta=0$, we set $A_r=0$. Then
\begin{equation}
  \label{eq:intpsi}
  \begin{split}
    B_\theta&=-{\partial A_s\over\partial r}={1\over
      r}{\partial\Psi\over\partial\theta}\\
B_s&={1\over r}{\partial (rA_\theta)\over\partial
  r}={\partial\Psi\over\partial s}
  \end{split}
\end{equation}
and we can find $\vec{A}$ simply by integrating. This results in
\begin{equation}
  \label{eq:vecpol}
  \begin{split}
A_\theta=&{k'\over 8}r^3\sin 2\theta\\
A_s=&\left(-{k\over 2}r^2+{k''\over 48}r^4\right)\cos 2\theta
\end{split}\end{equation}
or in Cartesian:
\begin{equation}
  \label{eq:veccart}
\vec{A}=\left(-{k'\over 4}xy^2,{k'\over 4}x^2y,-{k\over 2}(x^2-y^2)+{k''\over 48}(x^4-y^4)\right)
\end{equation}

It is interesting to compare this with the VAD\cite{venturini1998map}
vector potential
\begin{equation}
\vec{A}=\left(\frac{k'}{4}\left(x^3-x y^2\right),\frac{k'}{4} \left(x^2 y-y^3\right),-\frac{k}{2} \left(x^2-y^2\right)+\frac{k''}{12}\left(x^4-y^4\right)\right)
\end{equation}

The two vector potentials differ by the gradient of the following function: 
\begin{equation}
\chi(x,y,s)=\frac{1}{16} \left(x^4-y^4\right) k'
\end{equation}

\subsection{Lorenz Gauge}
Neither of these two vector potentials satisfy $\nabla\cdot\vec{A}=0$,
the Lorenz gauge. We can add any multiple of $\chi$ and it turns out
that adding $\nabla\chi/3$ to eqn.\,\ref{eq:veccart} does satisfy Lorenz gauge to required order:
\begin{equation}
\vec{A}=\left(\frac{k'}{4}\left({x^3\over3}-xy^2\right),\frac{k'}{4}
  \left(x^2 y-{y^3\over 3}\right),-\frac{k}{2} \left(x^2-y^2\right)+\frac{k''}{24}\left(x^4-y^4\right)\right)
\end{equation}

\subsection{Magnetic Hamiltonian}
I choose to use eqn.\,\ref{eq:veccart} because it's the simplest. Then the Hamiltonian can be written:
\begin{equation}
  \label{magHz2}
\begin{split}
H&={k(x^2-y^2)\over 2}+{P_x^2\over 2}+{P_y^2\over 2}+\\
&+{(P_x^2+P_y^2)^2\over 8}+{k'xy(yP_x-xP_y)\over 4}-{k''(x^4-y^4)\over 48}
\end{split}
\end{equation}
The generating function which will eliminate derivatives of $k$ is
\begin{equation}
  \label{magGen}
  \begin{split}
    G(x,P_X,y,P_Y)=&xP_X+yP_Y+{k'\over 48}(x^4-y^4)+ \\
    &-{k\over 12}\left[(x^3+3xy^2)P_X-(3x^2y+y^3)P_Y\right],
  \end{split}
              \end{equation}
which, to the same order yields transformation
\begin{equation}
  \label{tram}
  \begin{split}
    x&=X+{k\over12}(X^3+3XY^2)\\
    P_x&=P_X-{k\over4}\left[(X^2+Y^2)P_X-2XYP_Y\right]+{k'\over12}X^3,
  \end{split}
\end{equation}
and similarly for $(y,P_y)$.  The transformed Hamiltonian is
\begin{equation}
\label{magH}
\begin{split}
  H^\ast=&{k\over 2}(X^2-Y^2)+{1\over 2}(P_X^2+P_Y^2)+\\
  &+{1\over 8}(P_X^2+P_Y^2)^2-{k\over 4}(X^2+Y^2)(P_X^2-P_Y^2)\\
  &+{k^2\over 12}(X^4+Y^4)+{k^2\over 2}X^2Y^2.
\end{split}
\end{equation}
Notice the similarity to eqn.\,\ref{Hstar}: in fact all terms are identical
except the last two, which only differ in their coefficients. 

Applying the same procedure as in the electrostatic case, we find
\begin{equation}
  \label{racm}
  \Delta P_x=-\int k^2ds\ \left({x^3\over 3}+xy^2\right)
\end{equation}

Or the fractional change in focusing strength:
\begin{equation}
  \label{fracm}
  {\Delta f_x\over f}={1\over fL}\left({x^2\over 3}+y^2\right)
\end{equation}
where $L$ is the effective length.

\clearpage
\bibliographystyle{unsrt} 
\bibliography{Baartman,BaartmanDN,Others}

\begin{thebibliography}{1}

\bibitem{Baartman:1997zza}
R.~Baartman.
\newblock {Intrinsic Third Order Aberrations in Electrostatic and Magnetic
  Quadrupoles}.
\newblock In {\em Proc.\ Particle Accelerator Conference, 12-16 May 1997,
  Vancouver, British Columbia, Canada}, pages 1415--1417. IEEE, 1997.

\bibitem{ruth1986single}
Ronald~D Ruth.
\newblock Single particle dynamics and nonlinear resonances in circular
  accelerators.
\newblock In {\em Lecture Notes in Physics}, pages 37--63. Springer, 1986.

\bibitem{matsuda1972third}
H~Matsuda and H~Wollnik.
\newblock Third order transfer matrices for the fringing field of magnetic and
  electrostatic quadrupole lenses.
\newblock {\em Nuclear Instruments and Methods}, 103(1):117--124, 1972.

\bibitem{baartman_9521}
R.~Baartman.
\newblock {Aberrations in Electrostatic Quadrupoles}.
\newblock Technical Report TRI-DN-95-21, TRIUMF, 1995.

\bibitem{venturini1998map}
M~Venturini, D~Abell, and A~Dragt.
\newblock Map computation from magnetic field data and application to the lhc
  high-gradient quadrupoles.

\end{thebibliography}
\end{document}